\documentclass[a4paper,american]{scrartcl}
\usepackage{lmodern}
\usepackage[utf8]{inputenc}
\usepackage[T1]{fontenc}
\usepackage{color}
\usepackage{babel}
\usepackage{amsmath}
\usepackage{amssymb}
\usepackage{setspace}
\usepackage[authoryear,round]{natbib}
\usepackage[font=small]{quoting}
\usepackage{latexsym}
\usepackage[bookmarks=false,pdfborder={0 0 0},colorlinks=false]{hyperref}

\makeatletter

\renewcommand{\vec}{\boldsymbol}

\renewcommand{\Im}{\operatorname{Im}}

\makeatother

\begin{document}

\title{The Physics and Metaphysics of Primitive Stuff}
\subtitle{forthcoming in the British Journal for the Philosophy of Science}

\author{Michael Esfeld%
\thanks{Université de Lausanne, Faculté des lettres, Section de philosophie,
1015 Lausanne, Switzerland.\ E-mail: \protect\href{mailto:Michael-Andreas.Esfeld@unil.ch}{Michael-Andreas.Esfeld@unil.ch}%
}, Dustin Lazarovici%
\thanks{LMU Munich, Mathematical Institute, Theresienstr.\ 39, 80333 Munich, Germany.\ E-mail: \protect\href{mailto:Lazarovici@math.lmu.de}{Lazarovici@math.lmu.de}%
}, Vincent Lam%
\thanks{Université de Lausanne, Faculté des lettres, Section de philosophie,
1015 Lausanne, Switzerland. School of History, Philosophy, Religion and Classics, The University of Queensland, 
St Lucia QLD 4072, Australia.\ E-mail: \protect\href{mailto:Vincent.Lam@unil.ch}{Vincent.Lam@unil.ch}%
}, Mario Hubert%
\thanks{Université de Lausanne, Faculté des lettres, Section de philosophie,
1015 Lausanne, Switzerland.\ E-mail: \protect\href{mailto:Mario.Hubert@unil.ch}{Mario.Hubert@unil.ch}%
}}
\maketitle
\begin{small}
\begin{abstract}
The paper sets out a primitive ontology of the natural world in terms of primitive stuff, that is, stuff that has as such no physical properties at all, but that is not a bare substratum either, being individuated by metrical relations. We focus on quantum physics and employ identity-based Bohmian mechanics to illustrate this view, but point out that it applies all over physics. Properties then enter into the picture exclusively through the role that they play for the dynamics of the primitive stuff. We show that such properties can be local (classical mechanics), as well as holistic (quantum mechanics), and discuss two metaphysical options to conceive them, namely Humeanism and modal realism in the guise of dispositionalism.

\medskip{}

\end{abstract}
\end{small}

\section{Introduction\label{sec:Introduction}}
There are two main options pursued in current
research on the ontology of quantum physics. One option is to
take the formalism of the quantum theory that one adopts to refer to the quantum state, represented by the universal wave-function, that is, the wave-function of the whole universe. Consequently, the quantum state is the physical object of the formalism
of quantum physics. This option is pursued in the quantum theory going
back to \citet{Everett:1957aa} (see also \citealt{Albert:1996aa}
and the papers in \citealt{Saunders:2010aa} and \citealt{Albert:2013aa}
for discussion). However, since the quantum state is defined on a very high-dimensional
space, namely the configuration space of the universe, this option implies that one is committed to that very
high-dimensional space being the space in which the fundamental physical
reality is situated.

If one shrinks back from that consequence and maintains that quantum physics is about matter existing in three-dimensional space or four-dimensional
space-time, one is committed to what is known as a \emph{primitive
ontology} as regards that matter. The role of the quantum state, represented
by the universal wave-function, then is limited to the dynamics, that is, its role
is to guide or govern the temporal development of the distribution
of matter in physical space (see \citealt{Allori:2008aa}). In brief, the motivation for the primitive ontology option is to uphold the commitment to physics being about matter in ordinary space also when it comes to quantum physics, although the quantum state is defined on a very high-dimensional space (see e.g.\ \citealt{Monton:2006aa}, \citealt{Maudlin:2010aa}, \citealt{Belot:2012aa}).

It is obvious that a dualism consisting in a conjunction of these
two options is not an attractive position: if one takes the quantum
state as represented by the universal wave-function to be the physical
object of quantum physics, then that state as it exists in the very
high-dimensional space on which the universal wave-function is defined,
is the physical reality. There then is no point to take that state
to be also the state of matter distributed in three-dimensional space. The task rather is to show how the
dynamics of the quantum state existing in the configuration space
of the universe can be such that this state develops in that space
\textendash{} e.g.\ through decoherence \textendash{} into something
that can account for our experience of matter being distributed in
a three-dimensional space.

By the same token, if one commits oneself to a primitive ontology
of matter distributed in three-dimensional space being the referent of the formalism of quantum physics,
then there is no point in adding to that commitment a commitment to
the quantum state existing in the high-dimensional configuration space
of the universe. The reason is, in brief, that it is not intelligible
how the quantum state could fulfill the role that it has in the primitive
ontology theories of quantum physics, namely to guide the temporal
development of the primitive ontology, if it were a physical object
on a par with the primitive ontology, but existing in another space;
it would, for instance, be unclear how a field existing in the very high-dimensional
configuration space of the universe, represented by the universal
wave-function, could guide the motion of matter in three-dimensional
space. Since the quantum state enters the primitive ontology theories
through the role that it plays for the temporal development of the
primitive ontology, it is reasonable to regard it as nomological, by
contrast to a physical entity on a par with the primitive ontology (cf.\ \citealt*[chs.\ 11.5 and 12]{Durr:2013aa}). We will consider in sections \ref{sec:Humean-best-system} and \ref{sec:Modal-realism} two proposals to spell out what it means that the quantum state is nomological, namely Humeanism and dispositionalism.

This paper is concerned with the second option. Its aim is to push
the idea of a primitive ontology of quantum physics to its ultimate
consequence and to show that the primitive ontology option applies
throughout physics. The ultimate
consequence is to maintain that matter is primitive stuff, \emph{materia
prima}, having as such no physical properties
at all. What is usually regarded as physical properties enters
into the theory through its role for the dynamics of the primitive
stuff, that is, through its nomological role. In other words, the
way in which the primitive ontology theories of quantum physics are
often presented, namely in terms of introducing the elements of
the primitive ontology as being characterized by classical properties
such as mass and charge, is incoherent, as is the dualism of a primitive
ontology existing in three-dimensional space and a quantum state existing
in configuration space. The reason for this incoherence is that the
dynamics of classical physics is fundamentally different: in classical
physics, dynamical
variables such as mass and charge are attributed to point particles
taken individually. Given the laws of classical physics, the distribution of mass and charge in the universe fixes how the particles move.

In quantum physics, by contrast, it is in
general not possible to attribute a wave-function to the particles
taken individually, but in the last resort only to the whole configuration
of matter in the universe at a given time. In the primitive ontology
theories of quantum physics, the wave-function then has the job to
fix the temporal development
of the configuration of matter (in a deterministic or probabilistic manner). It is incoherent to assume that the
determination of the dynamics encoded in the wave-function is superimposed
on a determination of the dynamics through the classical properties
of the particles taken individually, that is, their mass and their charge.
In brief, either the dynamics is determined
from above so to speak, namely by variables that apply only to
the primitive ontology as a whole, or the dynamics is determined from
below, namely by variables belonging to the elements of
the primitive ontology taken individually. Since the latter option
is excluded for quantum physics, it is reasonable to pursue the former
one. This implies taking the primitive ontology to be primitive
stuff, instead of particles that are equipped with intrinsic properties
each, and conceiving the dynamics as being determined by variables
belonging to the whole configuration of the primitive stuff.

In the next two sections, we first elaborate on the metaphysics and then on the physics of matter as primitive
stuff, using the formalism of Bohmian mechanics for identical particles.
In section \ref{sec:Humean-best-system}, we show how a recent proposal
for a Humean conception of the dynamical variables can shed light on this
view of matter. In section \ref{sec:Modal-realism},
we apply this proposal to the more ambitious metaphysical stance according to which the dynamical variables literally determine the temporal development
of an initial configuration of primitive stuff.

\section{Primitive Ontology: Primitive Stuff\label{sec:Primitive-ontology}}

The term \emph{primitive ontology} goes back to \citet*[ch.\ 2, see end of section 2.2, paper originally published 1992]{Durr:2013aa}.
They write:
\begin{quote}
What we regard as the obvious choice of primitive ontology\textemdash the
basic kinds of entities that are to be the building blocks of everything
else (except, of course, the wave function)\textemdash should by now
be clear: Particles, described by their positions in space, changing
with time\textemdash some of which, owing to the dynamical laws governing
their evolution, perhaps combine to form the familiar macroscopic
objects of daily experience. (Quoted from the reprint in \citealt*[p.\ 29]{Durr:2013aa};
a forerunner of this notion can be found in \citealt[p.\ 46]{Mundy:1989aa})
\end{quote}
This term has been created in the context of quantum mechanics in order to remind
us of the fact that the formalism of quantum mechanics is supposed
to represent something, namely matter in space, and is supposed to
describe its behavior, for instance in measurement situations. The
first sense in which the ontology of matter distributed in physical
space is primitive is that this ontology cannot be inferred from the
formalism of textbook quantum mechanics, but has to be put in as the
referent of that formalism. According to the proposal pursued in this paper, that ontology is
furthermore primitive in the sense that it consists in primitive stuff,
that is, stuff that has as such no physical properties. Dürr, Goldstein, and Zangh\`{i} allude to this meaning of \textquotedblleft primitive\textquotedblright{}
in the quotation above when they say that the particles are described
only by their position in space. That is to say, a particle being
located at a point of space merely signifies that the point in question
is occupied instead of being empty. But as far as the primitive ontology
is concerned, there are no physical properties \textendash{} such
as a mass or a charge \textendash{} instantiated by the particle.

The de Broglie-Bohm theory, going back to \citet{Broglie:1928aa}
and \citet{Bohm:1952aa} and known today as Bohmian mechanics (see \citealt*{Durr:2013aa})
is the oldest primitive ontology theory of quantum mechanics. Bohmian
mechanics puts forward a discrete primitive ontology of point particles,
whereby, as mentioned above, a particle being located at a point of
three-dimensional space means that the point in question is occupied
by primitive stuff instead of being empty. What accounts for the primitive
stuff occupying points being particles is that, according to Bohmian
mechanics, there are continuous lines of occupation of points in space-time,
so that there are worldlines constituting particle trajectories. In Bohmian
mechanics,
the role of the wave-function, developing according to the Schrödinger
equation, is to determine, via what is known as the guiding equation, the velocity of each particle at any time
$t$ given the position of all the particles at $t$. We will go into the physics of Bohmian mechanics in the next section. For present purposes, it is only important to note that velocity
is not a property that the particles have over and above being located
in space, but simply the first temporal derivative of position.

The view according to which all the physical properties, including mass and charge, are best understood at the level of the wave-function rather than at the level of the Bohmian particles has been suggested in the literature on the basis of experimental considerations involving interference phenomena, for instance in the context of the Aharonov-Bohm effect and of certain interferometry experiments (see e.g.\ \citealt{Brown:1995aa} and references therein; cf.\ also most recently \citealt{Pylkkanen:2014aa}). \cite{Brown:1996aa} explicitly discuss this view \textendash{} which they call the \emph{parsimonious view} \textendash{}, but only within the framework of a dualistic ontology that recognizes both the Bohminan particles and the wave-function as genuine ontological entities on their own right.  However, as we have argued above in section \ref{sec:Introduction}, there is no point in doing so; in particular, it remains entirely mysterious how the wave-function understood as a physical object on configuration space could guide the Bohmian particles. Indeed \citet[§ 4]{Brown:1996aa} acknowledge this fact when they concede that their \emph{parsimonious view} faces what they call the \emph{problem of recognition}, namely to explain how the wave-function of a given particle ``knows'' which particle to guide when there are several particle species in a region of overlap of the respective wave-functions, assuming a factorizable total wave-function for simplicity.

Furthermore, there are two primitive ontology theories of quantum
mechanics using the dynamics proposed by Ghirardi, Rimini, and Weber
(GRW) \citeyearpar{Ghirardi:1986aa}, which seeks to include the textbooks\textquoteright{}
postulate of the collapse of the wave-function upon measurement into
a modified Schrödinger equation. \citet[ch.\ 22]{Bell:2004aa} suggests that whenever there is a spontaneous localization
of the wave-function in configuration space, this development of the
wave-function in configuration space represents an event occurring
at a point in physical space. These point-events are today known as
\emph{flashes}; that term was introduced by \citet[p.\ 826]{Tumulka:2006aa}.
According to the GRW flash theory (GRWf), the flashes are all there
is in space-time. As far as the primitive ontology is concerned, the GRW flash theory
is the Bohmian particle ontology without the trajectories: instead
of particle trajectories \textendash{} that is, continuous lines of
occupation of points in space-time \textendash , there are only isolated
points being occupied by primitive stuff. 

Bohmian mechanics and the GRW flash theory both propose a primitive ontology of primitive stuff that is \emph{discrete}: particles or flash-events at space-time points. By contrast, Ghirardi, Grassi, and Benatti (1995) \nocite{Ghirardi:1995aa}
develop an ontology of a \emph{continuous} matter density distribution in
physical space (GRWm). The wave-function in configuration space and
its temporal development as described by the GRW equation represent
at any time the density of matter in physical space, and the spontaneous
localization of the wave-function in configuration space (its \textquotedblleft collapse\textquotedblright )
represents a spontaneous contraction of the matter density in physical
space, thus accounting for measurement outcomes and well localized
macrophysical objects in general (see also \citealt{Monton:2004aa}).
Again, matter is primitive stuff, as pointed out by \citet{Allori:2013aa}:
\begin{quote}
Moreover, the matter that we postulate in GRWm and whose density is
given by the $m$ function does not ipso facto have any such properties
as mass or charge; it can only assume various levels of density. (\citealt[pp.\ 331--332]{Allori:2013aa})
\end{quote}

Matter thus is gunk, filling all of space. This, however, implies that the primitive stuff admits
of degrees, as expressed by the $m$ function in the GRWm formalism:
there is more stuff at some points of space than at others, with the
density of matter at the points of space changing in time; otherwise,
the theory would not be able to accommodate variation. But it remains unclear what could constitute the difference in degrees of stuff at \textit{points} of space, if matter just is primitive stuff. The GRWm theory hence is committed to the view of matter being a bare substratum with its being a primitive fact that this substratum has various degrees of density at points of space or space-time. In other words, there is a primitive stuff-essence of matter that admits different degrees of density. On Bohmian
mechanics and the GRW flash ontology, by contrast, the only variation consists in some points
of space being occupied while others are empty, with there being a
change in time in which points of space are occupied. This ontology can then easily account for the concentration of matter in certain regions of space by maintaining that in some regions of space, more points are occupied than in other regions of space.

Nonetheless, Bohmian mechanics and the GRW flash theory face the following question: What is it that occupies points of space? In other words: What accounts for the difference between a point of space being occupied and its being empty? There are no intrinsic properties such as mass or charge available that could make up for that difference. That is to say, Bohmian particles or GRW flashes do not have an intrinsic essence constituted by intrinsic properties. Even if there are no intrinsic properties, one could still maintain that Bohmian particles or GRW flashes have a primitive thisness (haecceity). However, haecceitism is a very controversial metaphysical stance. In any case, it is a purely metaphysical view that is always available if one is willing to pay the price, physics be as it may. In other words, there is no motivation for haecceitism from physics (especially given the explicit and generalized permutation invariance that we will explain below in section \ref{sec:matter-as-primitive-stuff}). It seems hence that also in the case of a primitive ontology of discrete objects (particles, flashes), we have to fall back into admitting a primitive stuff-essence of matter that accounts for the difference between a point of space being empty and its being occupied. The only difference between a primitive ontology of discrete objects and a primitive ontology of gunk would then be that in the latter case that primitive stuff-essence also has to include different degrees of density at points of space. In a nutshell, it seems that the primitive ontology theories of quantum physics are committed to conceiving matter as a Lockean bare substratum.\footnote{We are grateful to one of the referees for raising this objection.}

This consequence puts these theories in an uncomfortable position: the commitment to a bare substratum is a controversial metaphysical stance. One may motivate this stance by claiming that there has to be a primitive stuff-essence at the bedrock of matter. But one 
can also with reason object that a primitive stuff-essence in the guise of a bare substratum is mysterious. In any case, again, the view that there is a primitive ontology of physics is well-motivated \textendash{} since physics, including quantum physics, can with good reason be taken to be about matter in space-time \textendash{}, but there is no motivation from physics to conceive the primitive ontology in terms of a primitive stuff-essence of matter (cf.\ the objection that \citealt[p.\ 136 note 15]{Ladyman:2007b}, raise against Bohmian particles). The upshot of these considerations hence is that if one admits an essence of matter, that essence be better constituted by properties \textendash{} or relations, as we shall argue \textendash{}, but never be primitive.

The impasse into which the question of what accounts for the difference between a point of space being occupied and its being empty runs is a consequence of conceiving the primitive ontology theories in terms of a commitment to absolute space into which matter is inserted. Only in the case of a dualism of there being points of space and matter occupying these points does that question arise. However, whereas working with an absolute background space certainly is an elegant manner of presenting these theories (at least as long as the issue of including gravity is left out), there is no reason why the primitive ontology theories should be commitment to a dualism of matter and space. In other words, speaking in terms of points of space being occupied or empty just is a convenient manner of
setting out the view of matter as primitive stuff, but should not be taken literally. If we conceive the primitive ontology in terms of discrete objects (particles, flashes), we can formulate its core claim in the following manner:  \textit{matter is primitive stuff. It is discrete, consisting in matter points. These are
matter points, because there is a non-vanishing
three-dimensional distance between any two such points.} In other words, they are matter points in virtue of being connected by metrical relations. In that way, switching from absolutism to relationalism about space removes the commitment to a bare substratum or a primitive stuff-essence because it opens up the possibility to conceive the primitive stuff in terms of standing in metrical relations that are its essence. 

A primitive ontology theory that treats
matter as primitive stuff, but seeks to avoid a commitment to a primitive stuff-essence or bare substratum cannot but adopt the Cartesian characterization of matter in terms
of spatial extension.
In a nutshell, what distinguishes points of a primitive matter stuff
from points of a hypothetical primitive mental stuff only is that the
former, by contrast to the latter, are connected by metrical relations. Nonetheless, there are no space-time points.
There are substances that are not extended in themselves (points).
These are material, because they are connected by spatial relations
and move, so that there is change in their spatial relations and thus
a temporal development of the spatial configuration of these point-substances. If they were not connected by spatial relations, but by hypothetical fundamental mental relations, they would not be primitive matter stuff (matter points) and not be physical entities, but primitive mental stuff.

Hence, what makes it that a point is a matter point is nothing intrinsic of that point \textendash{} no intrinsic properties, no primitive thisness, no bare substratum or primitive stuff-essence \textendash{}, but the fact that it stands in spatial relations. The view of matter as primitive stuff thereby joins the stance known as (moderate) ontic structural realism in claiming that the identity of the fundamental physical objects, namely the matter points in this case, is provided by certain relations, namely metrical relations (see \citealt{Esfeld:2008aa}). Nonetheless, these relations are strong enough to allow the matter points to fulfill Leibniz' principle of the identity of indiscernibles in that they can be absolutely discernible:\footnote{Recall that two entities are absolutely discernible if and only if there is a physically meaningful monadic predicate or, more generally, a physically meaningful formula with one free variable that applies to one but not to the other.} it is possible that each matter point is distinct from all the other ones by some of the distance relations that it bears to other matter points. If they are absolutely discernible, the matter points are individuals by the standards commonly used in the philosophy of physics. As the development of ontic structural realism has made clear, neither individuality nor absolute distinguishability need to be grounded in intrinsic features (\citealt{Ladyman:2007aa} nicely illustrates this point). In a nutshell, the famous slogan ``No entity without identity'' coined by \citet[p.\ 23]{Quine:1969aa} applies to the matter points, although they are primitive stuff: they do not have an intrinsic identity, but a relational one provided by spatial distances that can be so strong that it makes them absolutely discernible entities. Furthermore, if they are particles, their trajectories endow them with a diachronic relational identity: each particle is absolutely discernible from the other ones not only by its position at any given time, but also by its history.

To sum up, the matter points are primitive stuff in the following two senses: (a) they are fundamental \textendash{} that is, they are not composed of anything else, but their configurations compose everything else; (b) they are primitive objects \textendash{} that is, they do not have an intrinsic essence constituted by intrinsic properties. However, they are not primitive in the sense of possessing a primitive stuff-essence: they are not bare substrata. If they have an essence, the relations in which they stand and that individuate them, namely the metrical relations, are their essence.

As regards the metaphysical literature on objects, the primitive ontology of primitive stuff motivated by quantum physics falls into neither of the two main stances: it obviously does not conceive objects as bundles of properties, since there are no such properties available in quantum physics; and it does not conceive objects as bare substrata either. By contrast, it joins the stance of moderate ontic structural realism in the philosophy of science by being committed to objects, but maintaining that these objects are individuated by certain relations in which they stand, namely metrical relations. Coming back to the metaphysical literature, the view that comes closest to this one is the proposal by \cite{Heil:2003aa,Heil:2012aa} according to which there are substances, but these substances are not bare particulars: they always exist in certain ways (modes) of being that individuate them. However, whereas Heil conceives these ways of being as intrinsic features of these substances and refuses to admit any relations on the ontological ground floor (see \citealt[ch.\ 7]{Heil:2012aa}), in the primitive ontology of quantum physics, there are, as mentioned above, no such intrinsic features available. We therefore have to go structural, conceiving the relations in which these substances stand as their basic way of being, namely the metrical relations. Nonetheless, we thereby join an old tradition, namely the Cartesian one of conceiving matter as \textit{res extensa} only.

\section{The Physics of Matter as Primitive Stuff}\label{sec:matter-as-primitive-stuff}
Let us turn to Bohmian mechanics in order to illustrate the physics of matter as primitive stuff, since Bohmian mechanics is the best known example of a primitive ontology formulation of non-relativistic quantum mechanics and since it is the only primitive ontology theory for which there is a version worked out in terms of permutation invariance available. Although, if spelled out in a consequent manner, the primitive ontology of matter as primitive stuff should go with relationalism about space, we will use for the sake of this illustration the formulation in terms of an absolute background space with some points of that space being occupied, whereby these occupied points make up the configuration of matter in the universe. Our primary aim in this section is to show what the physics of primitive stuff in contrast to the physics of material objects with intrinsic essences looks like. Casting that physics at the same time in relationalist terms about space and time would by far go beyond a single paper -- the main challenge in this respect is to investigate whether Bohmian mechanics admits a universal wave-function that has all the right symmetries to depend only on the metrical relations between the particles.

Bohmian mechanics, as commonly presented (see the papers in Dürr, Goldstein and Zanghì 2013 and the textbook Dürr and Teufel 2009), is a theory about point particles moving in three-dimensional  space, whereby the quantum wave-function figures in a non-local law of motion for the configuration of particles. 
Usually, the theory is introduced by formulating the laws of motion on the configuration space $\mathbb{R}^{3N}$, where $N$ is the number of particles and $Q(t)=\left(Q_{1}(t),\dots,Q_{N}(t)\right) \in \mathbb{R}^{3N}$ represents their positions at time $t$. 

The configuration then evolves according to the guiding equation 
\begin{equation}\label{eq:standard-guiding-equation}
\frac{\mathrm{d}Q_k}{\mathrm{d}t}= \frac{\hbar}{m_k}\frac{\psi^*\nabla \psi}{\psi^* \psi} (Q_1,\dots , Q_N),
\end{equation}
where $\psi(q_1,\dots , q_n)$ is the wave-function representing the quantum state of the system. The time-evolution of this wave-function, in turn, is given by the  Schrödinger equation
\begin{equation}
\imath\hbar\frac{\partial\psi}{\partial t}= \Bigl(-\sum_{j=1}^N \frac{\hbar^2}{2m_j} \Delta_j + V(q_1,\dots , q_n) \Bigr)\, \psi,\label{eq:Schroedinger-equation}
\end{equation}
familiar from standard quantum mechanics. The non-local character of the law is manifested in the fact that the velocity of any particle at time $t$ depends on the position of every other particle at time $t$;  the law of motion, in other words, describes the evolution of the particle configuration \textit{as a whole}. This is necessary in order to take quantum non-locality \textendash{} as illustrated for instance by Bell's theorem (Bell 1987, ch.\ 2) \textendash{} into account.

The parameters $m_k$ appearing in equation \eqref{eq:standard-guiding-equation} and \eqref{eq:Schroedinger-equation} correspond to the mass of the $k$-th particle. Furthermore, we observe that for (static) electromagnetic interactions, the charges $e_k$ of the particles enter the Schrödinger equation via the Coulomb potential\footnote{If the full electromagnetic interactions are taken into account, the gradient in equation \eqref{eq:standard-guiding-equation} is replaced by a covariant derivative, into which the vector potential and the particle charges enter.}
\begin{equation} V(q_1,\dots , q_n) = \sum\limits_{i<j} \frac{e_i e_j}{\lVert q_i - q_j \rVert}.\end{equation}
We will address the status of these parameters later in this section. 

It is important to note that, on a fundamental level, there is only \textit{one} wave-function in Bohmian mechanics: the universal wave-function $\Psi$, guiding the motion of all the particles in the universe together. Nonetheless, in many relevant cases, it is possible to provide a description of a (suitably isolated) subsystem as an autonomous Bohmian system in terms of an \textit{effective} wave-function $\psi$, which is derived from capital $\Psi$ and the actual spatial configuration of the environment, that is,\ the rest of the universe that is ``ignored'' in the description of the subsystem. These effective wave-functions can be seen as the Bohmian analogue of the usual quantum wave-functions familiar from textbook quantum mechanics.

An easy mistake in connection with Bohmian mechanics is to confuse the theories' commitment to particle positions with a realism regarding any of the physical quantities commonly associated with quantum mechanical observables. In fact, the opposite is correct. Bohmian mechanics is a theory about the motion of particles, conceived as the basic constituents of matter, and hence a theory about the distribution of matter in space and time. A statistical analysis of this theory, for situations corresponding to quantum measurements, then reproduces the outcome statistics of textbook quantum mechanics in terms of the effective wave-function (the quantum state) $\psi$ of the measured microscopic subsystem. The respective measurement outcomes, however, do not reflect any properties that the particles possess over and above their spatial configuration, but are shown to arise from their disposition of motion, which is encoded in $\psi$, upon interaction with a macroscopic measurement apparatus. 

A paradigmatic example is the Bohmian treatment of spin. In the token measurement of spin, let's say in the z-direction, a particle is sent through an inhomogeneous magnetic field (a Stern-Gerlach magnet) and then detected on a screen to see if it was deflected upwards or downwards, corresponding to the measurement outcome ``spin up'' or ``spin down'', respectively. How does Bohmian mechanics account for this experiment?

Consider a particle whose (effective) quantum state is described by a spinor-valued wave-function of the form
\begin{equation}\label{spinorwf} \psi(q) = \begin{pmatrix} \varphi_1(q) \\ \varphi_2(q)\end{pmatrix} = \left(\varphi_1(q) \binom{1}{0}+\varphi_2(q)\binom{0}{1}\right).\end{equation}
The Schrödinger time-evolution for $\psi$ in an inhomogeneous magnetic field\footnote{More precisely, $\psi$ is governed by the Pauli equation, which is the non-relativistic wave-equation describing particles with spin.} is such that the part of the wave-function corresponding to the upper spin-component is propagating in the positive z-direction, whereas the part of the wave-function corresponding to the lower spin-component is propagating in the negative z-direction. Consequently, the two spin-components of the initial wave-function \eqref{spinorwf} become spatially separated. The trajectory of the particle, determined by equation \eqref{eq:standard-guiding-equation}, will then follow one of the two wave-packets, depending on its initial position, and thus hit the screen above or below the zero-line, corresponding to a measurement of ``z-spin up'' or ``z-spin down'', respectively (for a detailed account of spin in Bohmian mechanics see \citealt[ch.\ 8.4]{Durr:2009fk}, and \citealt{Norsen:2014aa}). Hence, we see in particular that the measured spin-value does \textit{not} correspond to any property that the particle possesses over and above its position. For a Bohmian particle to have ``spin up'' or ``spin down'' means nothing more and nothing less than to be ``guided'' by the part of the wave-function that corresponds to the upper or lower spinor-component, that is, to \textit{move} -- in the pertinent measurement-context -- in the respective way.
 
A different issue is the status of the dynamical parameters \textit{mass} and \textit{charge}, which occur in Schrödinger's equation \eqref{eq:Schroedinger-equation} and -- in the case of mass -- in the Bohmian guiding equation \eqref{eq:standard-guiding-equation}. The crucial observation here is that the way in which the parameters $m_k$ and $e_k$ figure in equation \eqref{eq:Schroedinger-equation}, and thus the evolution of the wave-function on configuration space, is insensitive to the actual configuration of the particles in physical space. For this reason, it is inappropriate, in general, to think of mass and charge as intrinsic properties of the Bohmian particles which are ``carried along'' as they move in space.

This point is illustrated very clearly in the following experiment: consider a charged particle whose (effective) wave-function is of the form $\psi = \phi_A + \phi_B$, where $\phi_A$ and $\phi_B$ are of equal size and shape but concentrated on two distant regions of space that we denote by $A$ and $B$, respectively. (Those regions could be surrounded by infinite high potential walls -- or, more simply put, a box -- to keep the wave-function from spreading.) The particle will be located in one of those regions, let's say in $A$. Not surprisingly, the trajectory of a second charged particle passing near $A$ will be affected by the electromagnetic interactions and deflected towards $A$, if it has opposite charge, or away from $A$ if it has equal charge as particle one. However, if that second particle were passing near region $B$, it would be affected \textit{in the very same way}, no matter how far away that is from the actual position of the other particle. This scenario demonstrates, firstly, the explicitly non-local character of Bohmian mechanics. It also shows that it would thus be wrong to think of charge in the familiar way as something localized at the position of the particles. A similar reasoning would apply to the particle mass, in so far as gravitational interactions play a role in quantum mechanics. 

A common reply to this issue is that mass and charge should be regarded not as properties of the particle, but as properties of the wave-function, the intuition (presumably) being that the absolute value of $\psi$ -- or rather $\lvert\psi\rvert^2$ -- can represent (something akin to) a charge distribution. But this view is untenable for a variety of reasons. To begin with, we have already seen that the view of (effective) wave-functions as physical entities over and above the particles is unwarranted. If the wave-function associated with a particle is not a physical entity, it cannot carry physical properties. Moreover, as soon as we consider an entangled wave-function of two or more particles, it will correspond to a (non factorizing) function on a high-dimensional configuration space and cannot be taken to represent a distribution of physical quantitates in three-dimensional space.  

So what attitude shall we adopt vis-à-vis mass and charge in Bohmian mechanics? First and foremost, one should take the theory seriously in its own right and acknowledge that all the (classical) intuitions that we associate with mass and charge are \textit{a priori} questionable. In the first instance, $m_1,\dots,m_N$ and $e_1,\dots, e_N$ are merely numerical parameters that appear in the formulation of the Bohmian laws of motion. To illustrate this point, let us assume for the moment that there exists but a single species of particles, i.e.\ that $m_k = m_l = m$ and $e_k = e_l = e$  for all $k,l \in \lbrace 1,\dots,N\rbrace$. Hence, we see from equations \eqref{eq:standard-guiding-equation} and \eqref{eq:Schroedinger-equation} that we are left with the same three constants $\hbar$, $m$, and $e$ appearing in the equation of motion for any of the $N$ particles. There is then plainly no reason to treat $m$ and $e$ differently from Planck's constant $\hbar$. In particular, there is no justification to attach $m$ or $e$ to the particles or to interpret them as localized physical quantities, any more than we would do for $\hbar$. Rather we would regard $m$ and $e$ as nothing more than two additional constants of nature, numerical parameters entering the equations of motion without referring to anything in the physical ontology.

However, the commitment to a single type of particle, that is, a single elementary mass and charge, is clearly unsustainable from a physical point of view. Modern particle physics introduces an entire zoo of elementary particles varying in mass or charge or both: electrons, positrons, muons, anti-muons, the nucleons, respectively their constituent quarks, and so on. Hence, there must be something in the world which makes it the case that certain terms (respectively certain coordinates) in the equations of motion refer to, say, an electron rather than a muon. 

\citet{Goldstein:2005b, Goldstein:2005a} demonstrated the possibility to account for the different species of elementary particles in modern particle physics by reformulating Bohmian mechanics in a way that reflects the ontological commitment to propertyless particles as primitive stuff, treating all particles as \emph{identical}. To appreciate what this means and how the reformulation is carried out, let us begin with the following observation. If we insist that particles are distinguished only by their position, that is, spatial relations instead of intrinsic properties, we note that the configuration space $\mathbb{R}^{3N}$ has too much mathematical structure in that it ``cares'' about permutations of the particle labels. That is to say the following: the nature of the Bohmian law of motion (being a first-order differential equation on configuration space) is such that it determines at every time $t$ the change of the system's spatial configuration depending on the current configuration $Q(t)$. However, unless one presupposes a primitive identity or haecceity of the particles, the instantaneous configuration of an $N$-particle system is completely characterized by a set of $N$ points in physical space that are designated as being occupied by matter. There are no intrinsic properties, nor internal or external relations distinguishing the configuration represented by the tuple $(Q_1, Q_2, \dots, Q_N)$ from, let's say, the configuration represented by the tuple $(Q_2, Q_1, \dots, Q_N)$ with the particles $1$ and $2$ interchanged. It is thus understood that  -- for so-called \emph{identical} or \emph{indistinguishable} particles --  the natural configuration space of an $N$-particle system is not $\mathbb{R}^{3N}$, but 
\begin{equation}
^{N}\mathbb{R}^{3}:=\left\{ S\subseteq\mathbb{R}^{3}\mid\sharp S=N\right\} ,\label{eq:natural-configuration-space}
\end{equation}
which is the set of all subsets of $\mathbb{R}^{3}$ containing exactly $N$ elements.
Note that this space lacks the mathematical structure to represent permutations of the particle labels in contrast to $\mathbb{R}^{3N}$: a point $Q(t) = \lbrace Q_1, Q_2, \dots, Q_N \rbrace \in \,^{N}\mathbb{R}^{3}$ -- in contrast to the ordered N-tupel $(Q_1, Q_2, \dots, Q_N)$ -- describes the fact that at time $t$ there is \emph{a} particle occupying space-point $Q_1$, \emph{a} particle occupying space-point $Q_2$, and so on; it does not state that particle $1$ occupies $Q_1$, particle $2$ occupies $Q_2$, etc. 

Consequently, the wave-function of the system should now be defined on the configuration space $^{N}\mathbb{R}^{3}$ as well, which in fact can be done (\citealt[section 4]{Goldstein:2005b}). Nevertheless, it is still more convenient, in general, to represent the quantum state as a function on $\mathbb{R}^{3N}$ (which can be regarded, mathematically, as the universal covering space of $^{N}\mathbb{R}^{3}$). As long as we consider a system in which all particles are associated with the same mass and charge, the demand of consistency then leads immediately to a wave-function that is symmetric or anti-symmetric under permutations of the particle coordinates and hence to the famous \textit{boson/fermion alternative}. In \citet{Durr:2006aa}, $^{N}\mathbb{R}^{3}$ was thus already introduced as the configuration space of  identical or indistinguishable particles, referring to a \textit{single} species of particles, and it is shown how the quantum statistics of identical particles thus arise in the Bohmian theory (see also \citealt[ch.\ 8.5]{Durr:2009fk}). 

However, we now note that as soon as we have to admit more than one value for the parameters $m_k$, the standard formulation of Bohmian mechanics breaks down. That is because equation \eqref{eq:standard-guiding-equation} no longer defines a law of motion on $^{N}\mathbb{R}^{3}$, since it discriminates different particles by their associated mass, while configurations represented on $^{N}\mathbb{R}^{3}$ do not do so. The basic idea of \citet{Goldstein:2005b, Goldstein:2005a} is thus to symmetrize equation \eqref{eq:standard-guiding-equation} in order to get a permutation invariant equation, because any permutation invariant equation on $\mathbb{R}^{3N}$ defines, in a canonical way, a law of motion on $^{N}\mathbb{R}^{3}$, the configuration space of \emph{identical} particles. In this way, they show that we can treat \textit{all} particles as identical, while still accounting for the empirical data that, as usual, are explained in terms of a particle ``zoo''. 

To preserve equivariance of the law, i.e.\ the conservation of total probability by the Bohmian flow, the symmetrization has to be done in the following way. The standard guiding equation \eqref{eq:standard-guiding-equation} can be written in the form 

\begin{equation}\label{eq:guidingeq-2}
\frac{\mathrm{d}Q}{\mathrm{d}t}=\frac{j\left(Q(t)\right)}{\rho\left(Q(t)\right)},
\end{equation}
where 
\[
\rho=\psi^{\ast}\psi
\]
is the probability density and $j=\left(\vec{j}_{1},\dots,\vec{j}_{N}\right)$ with 
\[
\vec{j}_{i}=\frac{\hbar}{m_{i}}\Im\psi^{\ast}\nabla_{i}\psi
\]
the probability current corresponding to the system's wave-function $\psi$. In equation \eqref{eq:guidingeq-2}, numerator and denominator have to be symmetrized independently by summing over all possible permutations of the particle labels $1,\dots, N$.  Hence, we get a new, permutation-invariant guiding equation, which reads 
\begin{equation}
\frac{\mathrm{d}Q_{k}}{\mathrm{d}t}=\frac{\sum_{\sigma\in S_{N}}\vec{j}_{\sigma(k)}\circ\sigma}{\sum_{\sigma\in S_{N}}\rho\circ\sigma}(Q(t)).\label{eq:symmetrized-guiding-equation}
\end{equation}
Here, the sum goes over all elements of the permutation group $S_{N}$,
and 
\[
\sigma Q:=\left(Q_{\sigma^{-1}(1)},\dots,Q_{\sigma^{-1}(N)}\right)
\]
means that every coordinate $Q_{i}$ is assigned a new index $Q_{\sigma^{-1}(i)}$, changing the order in the $N$-tupel.

In this theory, which \citet{Goldstein:2005b, Goldstein:2005a} dubbed \emph{identity-based} Bohmian mechanics, we do not attribute \textit{a priori} any mass to any specific particle. The law of motion merely determines $N$ trajectories for $N$ particles, and it is a characteristic \textit{of this law} that one of those trajectories happens to behave -- at least in the relevant circumstances -- like the trajectory of a particle with mass $m_1$, another like the trajectory of a particle with mass $m_2$, and so on, depending only on the (contingent) initial conditions of the system, respectively the universe.

To illustrate how this works, let us discuss an example given in \citet[section 3]{Goldstein:2005b}
that compares the standard formulation of Bohmian mechanics with the \emph{identity-based} version. Consider a two-particle universe consisting of an electron with mass $m_{e}$ and a muon with mass $m_{\mu}$. Suppose, for simplicity, that they are in a non-entangled state $\Psi(q_{1},q_{2})=\phi(q_{1})\chi(q_{2})$ (note that we could symmetrize this wave-function, though this would be redundant when plugged into the symmetrized guiding-equation). Then, the standard guiding law \eqref{eq:standard-guiding-equation} leads to the following equations of motion: 
\begin{equation}
\begin{aligned}\frac{\mathrm{d}Q_{1}}{\mathrm{d}t} & =\frac{\hbar}{m_{e}}\Im\frac{\nabla\phi(Q_{1})}{\phi(Q_{1})},\\[1.3ex]
\frac{\mathrm{d}Q_{2}}{\mathrm{d}t} & =\frac{\hbar}{m_{\mu}}\Im\frac{\nabla\chi(Q_{2})}{\chi(Q_{2})}.
\end{aligned}
\label{eq:electron-muon-standard}
\end{equation}
In contrast, the symmetrized guiding equation (\ref{eq:symmetrized-guiding-equation}) reads
\begin{equation}
\begin{aligned}\frac{\mathrm{d}Q_{1}}{\mathrm{d}t} & =\frac{\frac{\hbar}{m_{e}}\left|\chi(Q_{2})\right|^{2}\Im\left(\phi^{\ast}(Q_{1})\nabla\phi(Q_{1})\right)+\frac{\hbar}{m_{\mu}}\left|\phi(Q_{2})\right|^{2}\Im\left(\chi^{\ast}(Q_{1})\nabla\chi(Q_{1})\right)}{\left|\phi(Q_{1})\right|^{2}\left|\chi(Q_{2})\right|^{2}+\left|\phi(Q_{2})\right|^{2}\left|\chi(Q_{1})\right|^{2}}\\[1.3ex]
\frac{\mathrm{d}Q_{2}}{\mathrm{d}t} & =\frac{\frac{\hbar}{m_{\mu}}\left|\phi(Q_{1})\right|^{2}\Im\left(\chi^{\ast}(Q_{2})\nabla\chi(Q_{2})\right)+\frac{\hbar}{m_{e}}\left|\chi(Q_{1})\right|^{2}\Im\left(\phi^{\ast}(Q_{2})\nabla\phi(Q_{2})\right)}{\left|\phi(Q_{1})\right|^{2}\left|\chi(Q_{2})\right|^{2}+\left|\phi(Q_{2})\right|^{2}\left|\chi(Q_{1})\right|^{2}}.
\end{aligned}
\label{eq:electron-muon-symmetrized}
\end{equation}

\noindent We see that equation (\ref{eq:electron-muon-standard}) ascribes -- or presupposes -- an intrinsic mass and thus a distinct type to every particle: particle $1$, described by the coordinates $Q_{1}$, is the electron with mass $m_{e}$, while particle $2$, described by the coordinates $Q_{2}$, is the muon with mass $m_{\mu}$. In equation (\ref{eq:electron-muon-symmetrized}), by contrast, neither $Q_{1}$ nor $Q_{2}$ is designated as the position of the electron, respectively the muon. \textit{A priori}, the two particles are distinguished only by the position that they occupy at time $t$. However, if we consider a situation in which $\phi$ and $\chi$ have disjoint support, say, when one wave-packet is propagating to the left and the other one to the right, one of the two sums in the nominators and denominators will be zero, so that the equation of motion \textit{effectively} reduces to equation (\ref{eq:electron-muon-standard}) (possibly with the indices 1 and 2 interchanged). This is to say, in particular, that in situations where the two-particle wave-function is suitably decohered, one of the particles will play the role of the electron -- being effectively described by equations \eqref{eq:standard-guiding-equation} and \eqref{eq:Schroedinger-equation} with the parameter $m_{e}$ -- while the other one will play the role of the muon -- being effectively described by equations \eqref{eq:standard-guiding-equation} and \eqref{eq:Schroedinger-equation} with parameter $m_{\mu}$. 

Which trajectory turns out to be guided by which part of the wave-function thereby depends only on the law of motion and the (contingent) initial conditions of the system, rather than on intrinsic properties of the particles. In fact, if both parts of the wave-function were brought back together and then separated again, one and the same particle could switch its role from being the electron to being the muon, and vice versa. Hence, like a particle's spin, we must conclude that to be an electron, a muon, or a positron, etc.\ is nothing more and nothing less than to \textit{move} -- in the relevant circumstances -- electronwise, muonwise, or  positronwise, and so forth. There are no properties in this theory defining different species of particles, but only \textit{primitive stuff}, following a law of motion that accounts for the phenomena conventionally attributed to a multiplicity of particle-types. 

Apart from such circumstances in which the different parts of the wave-function are well separated, one could say that the particles in the previous example are guided by a superposition of (what one would usually call) an electron wave-function and a muon wave-function. However, it would be misleading to claim that this amounts to a superposition of \textit{being} an electron and \textit{being} a muon. Ontologically, there are no superpositions of anything, only propertyless particles moving on definite trajectories. Rather, the labels ``electron'', ``muon'', etc. are meaningless in the general case. 

One obvious objection to the move proposed by Goldstein et. al. is that the guiding law \eqref{eq:symmetrized-guiding-equation} is much more contrived than the one in standard Bohmian mechanics. To some extent, this is a correct observation and we are indeed trading a sparse ontology for a more complicated mathematical formalism by endorsing the symmetrized theory. That notwithstanding, a few things can be said to address this worry. First, one should note that the apparent complexity of equation \eqref{eq:symmetrized-guiding-equation} is really just the price for expressing a law of motion for configurations in $^{N}\mathbb{R}^{3}$ on the coordinate space $\mathbb{R}^{3N}$ and doesn't necessarily amount to more complicated physics. Second, it should be noted that (modulo some subtleties discussed by \citealt{Goldstein:2005b,Goldstein:2005a}) the symmetrized theory will give rise to the familiar statistical description of subsystems in terms of effective wave-functions, which is really all that matters for most practical purposes. 

In this context, it should also be noted that, given the universal wave-function, the ``right'' statistical description of subsystems \textendash{} that is, the one agreeing with the predictions of standard quantum mechanics, arises for \textit{typical} initial conditions in terms of the particle configuration, that is, in \textit{quantum equilibrium} (see \citealt*[ch.\ 2]{Durr:2013aa}). Hence, the emergence of different particle types as empirically observed in nature  is not attributed to special initial conditions (quite the opposite), though it is ascribed to the particular form of the universal wave-function, i.e.\ to the physical law, if the latter is understood as nomological (we will expand on the nomological view of the wave-function in the upcoming sections). 

Finally, concerning the (empirical) content of the proposed theory, it should be emphasized that the trajectories described by \emph{identity-based} Bohmian mechanics will in general differ from those obtained from standard Bohmian mechanics, but that the statistical predictions for experimental outcomes are the same. In this sense, the symmetrized theory is \textit{empirically equivalent} to Bohmian mechanics and hence empirically equivalent to standard quantum mechanics. This shows, once more, that the physical ontology can neither be empirically determined, nor read off from the measurement-formalism of standard quantum mechanics, while, on the other hand, the choice of a primitive ontology \textit{can} supplement or enlighten the structure and formulation of the theory.

In particular, if the physical ontology is one of propertyless particles, this strongly suggests permutation invariant laws of motion in which all particles are treated as \emph{identical}  -- to borrow once more the terminology commonly employed in physics. Of course, this terminology is misleading in the sense that there is obviously a plurality of particles instead of just one particle. The meaning of \emph{identity-based} Bohmian mechanics -- and more particularly the meaning of permutation invariance within this framework -- is rather that we are committed to an ontology of primitive stuff in the sense of particles that do not possess any intrinsic properties nor any intrinsic identity. Permutation invariance thus means precisely that there is nothing to the particles beyond their position in the total configuration, in particular nothing that the laws of motion could refer to in order to establish a different dynamical role for different particles depending on some intrinsic characteristics.

To sum up, \emph{identity-based} Bohmian mechanics provides for a clear ontological meaning of permutation invariance: it encodes a primitive stuff ontology of individuals without any intrinsic identity and properties, though (absolutely) discernible in virtue of their position in the total configuration. Furthermore, permutation invariance applies here to all the particles, since there is only primitive stuff and no different species of particles, by contrast to concerning only the particles of the same species as in standard Bohmian mechanics or the wave-function supposedly corresponding to particles of the same species as in textbook quantum mechanics.

\section{The Humean Best System Analysis of the Dynamical Variables\label{sec:Humean-best-system}}

Staying within the framework of a primitive ontology of particles
as in Bohmian mechanics, how are we to conceive dynamical variables
such as mass or charge that are attributed to the particles taken
individually without making up for intrinsic essences that constitute different species of particles? Moreover, as mentioned in section \ref{sec:Introduction}, in the primitive ontology approach, it is reasonable to conceive the quantum state as a nomological entity by contrast to a physical entity on a par with the primitive ontology. But what does this mean? In this section and the next one, we will show that the main philosophical views about laws of nature can be employed in order to answer these questions. We will focus on Humeanism on the one hand and dispositionalism on the other.

Let us start with Humeanism. According to this view, the world is a vast mosaic of local matters
of particular fact, such as point particles being connected
only by relations of spatio-temporal distance. Given an initial configuration
of such point particles, there is nothing about that configuration
that puts a constraint on its temporal development. A certain temporal
development just happens to occur; there is nothing that guides, governs,
or determines it. Nevertheless, considering that temporal development
as a whole \textendash{} that is, the distribution of
the point particles throughout the whole of space-time
\textendash , that distribution exhibits certain patterns or regularities.
Consequently, if one sets out to put forward a description of the
distribution of the point particles in space-time,
one can do better than dressing a very long list that registers each
particle position. According to what is known as the Humean
best system account, the laws of nature are the axioms of the system
that achieves the best balance between being simple and being informative
in describing the distribution of matter throughout the whole of space
and time. In brief, laws of nature both simplify and are informative, striking the best balance between these two virtues
(see notably \citealt[ch.\ 3.3, pp.\ 72--75]{Lewis:1973aa}, and \citeyear[section 3]{Lewis:1994aa},
as well as \citealt{Cohen:2009aa}; there is no space here and it is not the aim of this paper to consider the internal problems of Humeanism). 

\citet[§ 5.2]{Hall:2009aa}, in particular, has put forward
a version of Humeanism that regards the vast mosaic
of local matters of particular fact as consisting
only in point particles standing in relations of spatio-temporal distance.
These particles are just primitive
stuff. Their distribution throughout space-time \textendash{} that is, the development of the metrical relations among the particles \textendash{} 
exhibits certain regularities. Suppose that the laws of classical
mechanics and electrodynamics figure in the Humean best system that
captures these regularities. Then dynamical variables such as mass
and charge appear in these laws. On the basis of these laws being part of
the Humean best system, one can then attribute properties like mass
and charge to the particles. That is to say: predicates such as \textquotedblleft mass\textquotedblright{}
and \textquotedblleft charge\textquotedblright{} apply to the particles. However, these predicates do not represent properties that the particles have \textit{per se}, as something essential or intrinsic to them. They apply to the particles 
in virtue of the contingent fact that their motion throughout the
whole of space-time happens to manifest certain regularities. Hence, what makes the application of these predicates true is nothing over and above
the distribution of primitive stuff throughout space and time. Nonetheless, this is not instrumentalism: Humeanism, applied to the primitive ontology approach in physics, is the view that the primitive ontology is the \textit{entire} ontology. However, the primitive ontology is an \textit{ontology} that stands on its own feet: it consists in theoretical entities such as point particles that exist in the world independently of observers and their beliefs.

This idea can also be applied to the wave-function in any of the primitive
ontology theories of quantum physics, notably Bohmian mechanics (see \citealt{Miller:2013aa},
\citealt{Esfeld:2014aa}, \citealt{Callender:2014aa}; see also already \citealt{Dickson:2000aa}). It can thus be employed to spell out what it means that the wave-function is nomological by contrast to being a physical entity on a par with the primitive ontology. Again,
the Humean mosaic
consists in the distribution of primitive stuff throughout the whole
of space-time \textendash{} such as particle trajectories, flash-events, or a matter density field. That distribution exhibits certain regularities. Suppose
that the laws of quantum mechanics figure in the Humean best system
that captures these regularities, and let these laws be the Bohmian
guiding equation and the Schrödinger equation, or a GRW-type equation
and a law establishing a link with the primitive ontology. Then a universal
wave-function describing the quantum state of the primitive stuff
appears in these laws, and the quantum state includes parameters such as mass and charge.
However, as these latter parameters do not require an ontological commitment
to anything more than the distribution of primitive stuff throughout
the whole of space-time in classical mechanics, so the quantum state is no addition to being:
given the whole distribution of the primitive stuff throughout space-time,
a law describing the temporal development of a universal wave-function
enters into the Humean best system as a means to achieve a description
of the distribution of the primitive stuff that strikes the best balance
between being simple and being informative about how the stuff is
distributed. This law simplifies and is informative in any case, since in a deterministic theory such as Bohmian mechanics, specifying the particle configuration and the wave-function at any given time is sufficient to capture the particle configuration at any other time. Given the law in which the wave-function figures, one can then attribute a quantum state
as represented by the universal wave-function to the particle configuration, in the
sense that the propositions doing so are true; but their truth-maker
is the distribution of the primitive stuff throughout the whole of
space-time, and not a quantum state that exists over and above the particle configuration.

Finally, coming back to the link between the primitive ontology approach to physics and relationalism about space-time, one can apply the Humean view of the primitive ontology being the entire ontology to space-time itself. Suppose that there is an initial configuration
of matter points that are primitive stuff and that are connected by
metrical relations. The matter points move so
that there is change in their spatial relations and thus a temporal
development of the initial configuration of matter points. On Humeanism,
there is nothing that puts a constraint on how that change has to
occur. Some such change just happens. If one combines Humeanism with
relationalism about space and time, there is furthermore nothing about
that initial configuration that singles out a particular motion as inertial
motion and a particular system of matter points as an inertial system. However,
as \citet{Huggett:2006aa} has shown, given the whole motion of the matter
points, there are some patterns or regularities in this motion that
make it possible to conceive a Humean best system achieving a good balance between
being simple and being informative in describing that motion. Based
on this best system, one can then single out a certain motion as inertial
and certain systems of matter points as inertial systems. One can thus account for absolute quantities such as acceleration in an ontology
of a Humean space-time relationalism applied to the laws of Newtonian
mechanics.

It is evident that this strategy can be put to work for any space-time,
not only a Newtonian one, as Humeanism is applicable to any primitive
ontology theory of matter. One has to assume an initial configuration
of extended stuff \textendash{} such as matter points being connected
by metrical relations \textendash ,
that configuration happens to develop in a certain matter. Considering
that development as a whole, it exhibits certain patterns or regularities.
Based on these patterns or regularities, there is a Humean best system
including the laws of both matter and space-time. Given that system,
dynamical variables can be attributed to the matter points, some systems
of them can be singled out as inertial systems, etc. On Humeanism,
whatever properties are attributed to matter or space-time come all
in one package, figuring in the Humean best system and being defined
by their role in that system, instead of being properties that
belong to matter or space-time as such.

\section{Modal Realism about the Dynamical Variables\label{sec:Modal-realism}}

Although Humeanism is a coherent philosophical way to conceive matter
as primitive stuff, showing how all the dynamical variables that are
commonly attributed to material objects can be derived from the Humean
best system, the physics of matter
as primitive stuff is not committed to the metaphysics of Humeanism. In other words, if the universal wave-function is nomological rather than a physical object on a par with the primitive ontology, its nomological character does not have to be spelled out in the framework of Humeanism about laws of nature. Indeed, there are many well known philosophical reservations against Humeanism in general. In particular, Humeanism cannot but regard it as a brute fact that the
regularities on which we rely in science as well as in everyday life
always
turn out to be well-confirmed. There is no constraint
at all on which local matters of particular fact can and which ones
cannot occur in the future of any given local matter of particular
fact, since what the laws of nature are depends on what there will
be in the future of any given local matter of particular fact, instead
of that future depending on the laws of nature. Hence, the laws of
nature cannot be invoked to answer the question of why certain regularities
\textendash{} such as e.g.\ those ones experienced as gravitation,
or those exhibited in the EPR-experiment \textendash{} always turn
out to be well-confirmed. There simply is no answer to that question
in Humeanism. Again, there is nothing incoherent about this position.
Yet the desire to obtain an answer to that question motivates the search for a more
ambitious metaphysical framework, that is, one that admits modal connections
in nature which put a constraint on what can and
what cannot happen in the universe given an arbitrary initial configuration
of matter.

The central anti-Humean answer to this question consists in anchoring
the laws of nature in properties that are attributed to the physical
systems (see notably \citealt{Bird:2007aa}). These properties are
such that it is essential for them to exercise a certain dynamical
role for the temporal development of the physical systems. The laws, in turn, express that dynamical role. 
These properties hence are \emph{dispositions} or \emph{powers}. Thus, on this view, mass and charge in classical mechanics are local properties
of the particles whose function is to accelerate the particles as described by the laws of Newtonian mechanics and classical electromagnetism. By way of consequence, on dispositionalism combined with a primitive ontology view of classical mechanics, the primitive stuff particles do indeed obtain properties each over and above standing in metrical relations. But these properties are not essential to the particles, and their role is not to provide an intrinsic identity of the particles; their job exclusively is a dynamical one, namely to put a constraint on how the particles move, given an initial configuration of particles whose identity is provided by the metrical relations in which they stand.

When it comes to quantum physics, it is no longer possible to conceive mass and charge as local dispositional properties or powers that belong to the particles taken individually, as various thought experiments such as the ones mentioned in section \ref{sec:matter-as-primitive-stuff} make clear (and see \citealt{Brown:1995aa} for more such experiments). Against this background, we have argued in section \ref{sec:matter-as-primitive-stuff} that what stands for mass and charge in the equation of motion are mere parameters without any direct ontological correlate. In brief, there are no mass and charge distributions influencing the motion of the particles. There only is the universal wave-function representing the quantum state. However, the quantum state is defined on configuration space. Hence, if one intends to attribute to the quantum state an ontological weight as a dynamical variable \textendash{} by contrast to regarding it simply as a convenient means to capture the salient regularities in the motion that the particles happen to take \textendash{}, one faces the difficulty of having to avoid the incoherent dualism mentioned at the beginning of this paper, namely the dualism of being committed to particles existing in physical space and a quantum state existing in configuration space. 

Dispositionalism avoids this pitfall in the following manner: as in classical physics the particles \emph{taken individually} instantiate dynamical, dispositional properties that determine their motion and that are represented by the mass and the charge variables in the laws of motion, so in quantum physics, the particle configuration \emph{as a whole} instantiates a dynamical, dispositional property that determines its temporal development and that is represented by the universal wave-function figuring in the laws of  (identity-based) Bohmian mechanics or the GRW theory. In brief, according to dispositionalism, the universal wave-function represents the common  disposition of motion of the particle configuration (see \citealt[pp.\ 77-80]{Belot:2012aa}, and \citealt[sections 4 and 5]{Esfeld:2014ab}). The motion of the particles is then such that, in certain specific circumstances, it is possible to consider them as if they were carrying  some local properties such as mass and charge that influence their motion, even though, from an ontological point of view, there is no such thing. Behaving like a ``massive particle'' or a ``charged particle'' is only the result of the particular particle motion (rather than its determinant) and contingent on the universal wave-function and the initial conditions.

In the framework of dispositionalism, the shift from classical to quantum mechanics hence amounts to a shift from local dynamical properties determining the motion of the particles to one holistic property of the particle configuration (the quantum state, represented by the universal wave-function) doing so. It is then more appropriate to characterize this property as a power than as a disposition, since if there is one holistic property of the particle configuration determining its temporal development, there is no question of an external stimulus or triggering condition for its manifestation (as a mass or a charge qua local property of a particle requires another massy or charged particle to manifest itself in the acceleration of the particles). In brief, on Bohmian dispositionalism, the primitive stuff particles collectively instantiate one power (represented by the universal wave-function) that determines their motion by determining their velocity. Consequently, this collective power relates strictly speaking all the particles with one another, determining their motion in tandem so to speak and thereby explaining quantum entanglement and the EPR correlations.

In general, if dispositions or powers instantiated by the particles taken individually can influence their motion, so can a collective power instantiated by the particle configuration. In both cases, the particle positions, consisting in the metrical relations that individuate them, by no means fix the disposition or power that determines the motion and thus the temporal development of the particle positions (i.e., their metrical relations). Such a disposition or power is in any case a modal property that has to be admitted in addition to the primitive ontology, but instantiated by the elements of the primitive ontology, as that what fixes what is possible and what is not possible about their motion. The metaphysical conception of dispositionalism applied to the primitive ontology of physics – that is, the commitment to properties as that what puts a constraint on the temporal development of the primitive stuff -- is the same in both cases. A holistic property or collective power doing so is no less intelligible and no more mysterious than local properties or powers doing so. The latter just are more familiar to us given our familiarity with classical physics and our unfamiliarity with quantum physics. In other words, the shift from local properties to holistic or collective properties is imposed upon us by the transition from classical to quantum physics. Any philosophical theory of properties has to adapt itself to this shift. Dispositionalism does so by countenancing dynamical properties (dispositions, powers) that are instantiated by the particle configuration as a whole instead of by the particles taken individually.

One can further illustrate this conception by linking it up with ontic structural realism. Since this collective power relates all the particles with one another, one can also conceive it as a structure defined on the configuration of the particles. This again is an ontic structure since, according to dispositionalism, it exists in the world over and above the primitive stuff (the particle configuration), albeit instantiated by it. However, one has to be careful not to confuse this ontic structure with the relational or structural individuation of the primitive stuff explained at the end of section \ref{sec:Primitive-ontology}: quantum entanglement conceived as an ontic structure in the framework of the physics of primitive stuff has nothing to do with the individuation or the discernibility of the physical entities; that individuation and discernibility, both at a time and in time, is obtained through the metrical relations in which the matter points stand at any time. It is not touched by the issue of Humeanism vs.\ modal realism (dispositionalism) as regards the laws of nature. The entanglement structure, by contrast, concerns only the dynamics of the matter points. If this structure exists over and above the matter points, it is a modal structure, putting a constraint on the temporal development of the matter points (their motion), whereas there is nothing modal about the metrical relations or structure insofar as they individuate the matter points, accounting for these points being matter points and being absolutely discernible. By way of consequence, only the modal realist but not the Humean is committed to the entanglement structure.

\section{Conclusion}

This paper started from recalling the two principled options for an ontology of quantum physics: (1) quantum state realism, according to which the quantum state as defined by the universal wave-function on configuration space is the physical reality, and (2) a primitive ontology theory, according to which the physical reality consists in matter existing in three-dimensional space or four-dimensional space-time. The aim of this paper was to push the primitive ontology option to its ultimate consequence, which is to regard matter as primitive stuff, namely as points that are matter points only in virtue of the metrical relations in which they stand; in particular, these matter points do not carry any intrinsic properties and do not possess any intrinsic identity. The metrical relations individuate them, making them (absolutely) discernible. These matter points are particles, if they persist and if their motion traces out continuous lines in space (worldlines that can be conceived as particle trajectories). Consequently, there are no different particle species in the fundamental ontology and ``permuting'' the particles obviously does not lead to any new physical situation. There just are propertyless particles qua matter points. We have shown how this view is naturally encoded in identity-based Bohmian mechanics.  

We then elaborated on two principled options for introducing physical properties through the role that they play for the dynamics of the primitive stuff. According to Humeanism, there is nothing over and above the primitive stuff throughout space and time. Given an initial configuration of primitive stuff, a certain temporal development of that configuration happens to occur. But there is nothing in nature that puts a constraint on which temporal development can happen and which one cannot happen. Given the distribution of primitive stuff throughout the whole of space–time, that distribution happens to exhibit certain patterns or regularities, which make it possible to formulate a Humean best system. The variables figuring in the Humean best system can then be attributed to the primitive stuff, but they do not represent an ontological commitment to anything over and above spatio-temporally extended primitive stuff.

According to modal realism, by contrast, there is something in nature over and above the primitive stuff that puts a constraint on its temporal development, fixing what can and what cannot happen given an initial configuration of primitive stuff. Dispositionalism spells this idea out in terms of dispositions or powers that the primitive stuff instantiates over and above being individuated by the metrical relations in which the matter points stand. These dispositions or powers enter the ontology only through the role that they play in determining a certain temporal development of the primitive stuff. They thereby ground the laws of nature. In classical physics, these are dispositions or powers that are instantiated by the matter points (the particles) taken individually; in quantum physics, there is in the fundamental ontology only one collective power instantiated by the configuration of the matter points (the particles) as a whole, tying their temporal development together.

To sum up, the primitive ontology option applies to both classical and quantum physics. If one endorses a primitive ontology of particles, the primitive ontology is the same in classical and quantum physics: particles as primitive stuff, individuated by the metrical relations in which they stand. The difference between classical and quantum physics concerns only the dynamics, namely the dynamical properties (dispositionalism) or predicates (Humeanism) attributed to the particles in order to account for the change in their metrical relations (that is, their motion): local properties in classical physics, a collective one in quantum physics.

Against the background of what has been achieved in this paper, we regard it as the foremost task for metaphysics to put the arguments for and against Humeanism and modal realism (dispositionalism) in the framework of a primitive ontology of primitive stuff shared by both these metaphysical stances. As concerns the physics, we take it to be the foremost task to elaborate on the link between the primitive ontology of primitive stuff and relationalism about space-time, thereby also extending this ontology beyond classical and quantum mechanics.

\section*{Acknowledgments}

We are grateful to three anonymous referees for helpful comments on the first version of the article. The work of Vincent Lam and Mario Hubert was supported by the Swiss National Science Foundation, grants no.\ PZ00P1\_142536 and PDFMP1\_132389 respectively.

\bibliographystyle{abbrvnat}
\bibliography{references_meta_final_2}

\end{document}